\documentstyle[twocolumn,aps,epsfig,graphics]{revtex}

\def\lsim{\lower.5ex\hbox{$\; \buildrel < \over \sim \;$}}
\def\gsim{\lower.5ex\hbox{$\; \buildrel > \over \sim \;$}}
\def\g{\ifmmode \gamma \else $\gamma$\fi}
\def\gs{\ifmmode \gamma \else $\gamma~$\fi}

\begin{document}

\title{Resonance Absorption and Regeneration in Relativistic Heavy Ion Collisions
}

\author{Sascha Vogel and Marcus Bleicher}

\address{Institut f\"ur Theoretische Physik,J.W. Goethe Universit\"at \\
Max-von-Laue-Str. 1, 60438 Frankfurt am Main, Germany } \maketitle

\noindent The regeneration of hadronic resonances is discussed for
heavy ion collisions at SPS and SIS-300 energies. The time
evolutions of $\Delta, \rho$ and $\phi$ resonances are investigated.
Special emphasize is put on resonance regeneration after chemical
freeze-out. The emission time spectra of experimentally detectable
resonances are explored. \vspace{.6cm}


The study of strongly decaying resonances is one of major topics that came
up in recent years \cite{pbmtalk}.
Recently, detailed experimental studies of hadron resonances became
available. Up to now $\rho, \Lambda^*, \Sigma^* \Delta, K^{0*}$ and $\Phi$ resonances
have been reconstructed from hadron correlations in heavy ion reactions at
the full SPS and highest RHIC energy
\cite{Markert:2004xx,Markert:2003rw,Markert:2002xi,Fachini:2003dx,Fachini:2003mc,Fachini:2004jx,Johnson:1999fv,Adler:2004zn}.\\
This exciting new data poses serious constraints on our
understanding and modeling of  hadronic interactions in the late
expansion stage of massive nuclear collisions. Especially the
apparent suppression of $K^{0*}$ and $\Lambda^*$ resonances (together
with $\Delta$ enhancement in central reactions compared to the
thermal model estimates \cite{Braun-Munzinger:2001ip} questions the notion of an instantaneous chemical
freeze-out as assumed in thermal models. In fact, it demonstrates
that the ideas of a directly observable Quark-Gluon-Plasma (QGP)
fireball without a longliving hadronic stage in scenarios with
strong supercooling \cite{Zabrodin:1998dk} seems not in line with
the experimental data. Whether hadronic statistical models can be
used to understand the resonance data is still not clear, see e.g.
\cite{Broniowski:2003ax,Torrieri:2004eb}\\
First ideas how to understand a modification of the resonance yields had
been brought forward by \cite{Johnson:1999fv} where a modification the inverse slope of the $\Phi$ meson
was discussed. Later this idea was
picked up by \cite{Torrieri:2001ue} in terms of a thermal models plus
absorption afterburner. However, this thermal absorption model fell short of the data, because it
neglected the regeneration of hadron resonances in the expansion stage. \\
Comprehensive studies of $\rho, K^{0*}, \Sigma, \Lambda$ and
$\Lambda^*$ hadron resonance yields and spectra within a transport
approach were presented in
\cite{Bleicher:2002dm,Bleicher:2002rx,Bleicher:2003ij} for SPS and
RHIC energies. For Pb+Pb collisions at 158 AGeV it was shown that
within the transport simulation a chemical freeze-out (where most of
the inelastic interactions cease and elastic and pseudo-elastic
interaction take over) around 6 fm/c is present. However, it was
also shown that elastic and pseudo-elastic reactions might still
alter the reconstructable resonance yields and
distributions due to absorption of daughter particles and regeneration (refeeding).\\

\begin{figure}[hbt]
\centerline{\epsfig {file=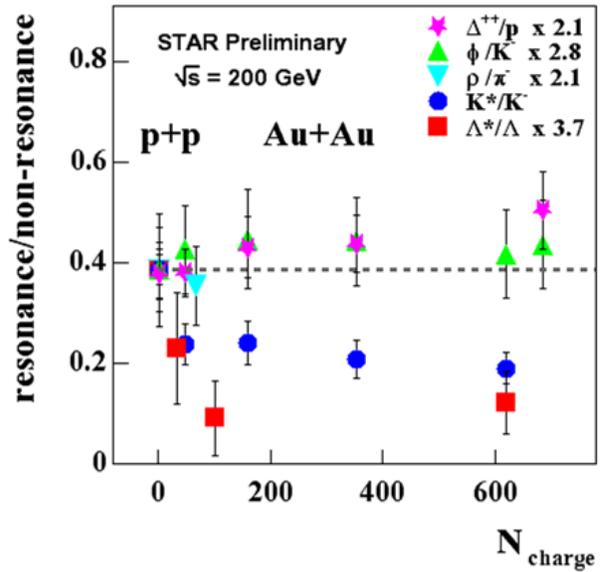,width=.5\textwidth,angle=0}}
\caption{ Compilation of measured particle ratios in pp and Au+Au
collisions at $\sqrt s=200$~AGeV as a function of centrality. Figure
is taken from \protect\cite{Markert:2004xx} } \label{f1}
\end{figure}

As shown in Fig. \ref{f1}, recent data by the STAR Collaboration
\cite{Markert:2004xx}  indicates interesting resonance to
groundstate ratios as a function of centrality. E.g., the $K^{0*} /
K$ and the $\Lambda^* / (\Lambda + \Sigma^0)$ ratios decrease, while
the $\Phi/K$ and $\Delta_{1232}^{++}/p$ ratios stay constant or
might even show a slight increase with centrality.

Let us now investigate possible mechanisms that produce at the same
time a slight increase of the $\Delta^{++}/p$ ratio and a
suppression of $K^*/K$ ratio. As a tool we employ the UrQMD model
\cite{Bass:1998ca,Bleicher:1999xi} which is based on the covariant
propagation of hadrons, resonances and strings.

\begin{figure}[hbt]
\centerline{\epsfig {file=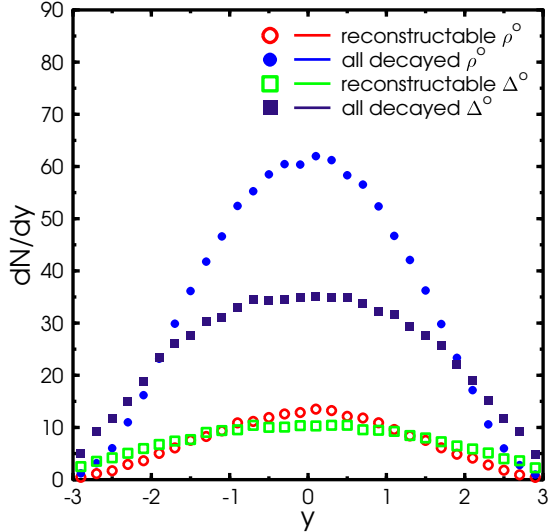,width=.5\textwidth,angle=0}}
\vspace{-10pt}
\caption{Rapidity distributions of $\Delta^0_{1232}$ and
$\rho^0_{770}$ for central Pb+Pb collisions at 158~AGeV. A strong
suppression of in hadron correlations reconstructable resonances
compared to those reconstructable via leptons (indicated as 'all
decayed') is visible. } \label{f2}
\end{figure}

\begin{figure}[hbt]
\centerline{\epsfig {file=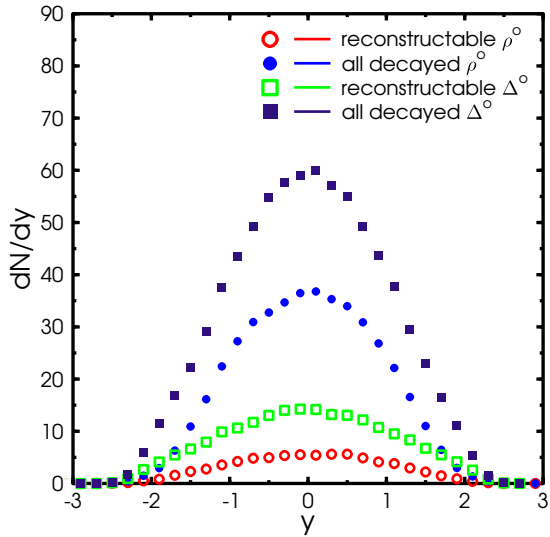,width=.5\textwidth,angle=0}}
\vspace{-10pt}

\caption{Rapidity distribution of $\Delta^0_{1232}$ and $\rho^0_{770}$
for central Pb+Pb collisions at 30 AGeV. A strong suppression of in
hadron correlations reconstructable resonances compared to those
reconstructable via leptons (indicated as 'all decayed') is visible.
} \label{f3}
\end{figure}

In the present calculations, modifications due to rescattering and
regeneration are present in nearly all spectra and yields of short
lived resonances. Especially for the $\Delta_{1232}$ and the
$\rho_{770}^0$, which both have lifetimes of approximately 1-1.5
fm/c this effect alters the spectra drastically. As will be shown
next the calculation predicts a difference between reconstructable
resonances and all decayed resonances. Fig. \ref{f2} shows the rapidity
distributions of the $\Delta^0_{1232}$ and the $\rho^0_{770}$ for
central Pb+Pb collision at the SPS energy of 158 AGeV. Full symbols
depict the resonances which decayed during the collision and, in
case of the $\rho$ are reconstructable in the dileptonic decay
channel. Open symbols show those resonances where the hadronic decay
products did not rescatter after their creation. In Fig. \ref{f2}
one observes a suppression by a factor of 6 for the $\rho$ meson and
a suppression by a factor of 3 for the $\Delta$. In Fig \ref{f3} one
observes a suppression by a factor of 7 for the $\rho$ and a
suppression by a factor of 6 for the $\Delta$. This can be
understood because of the different meson/baryon ratio at the
different energies, i.e. a higher baryon density at 30~AGeV compared
to 158~AGeV.

\vspace{-20pt}

\vspace{-35pt}

\begin{figure}[hbt]
\centerline{\epsfig
{file=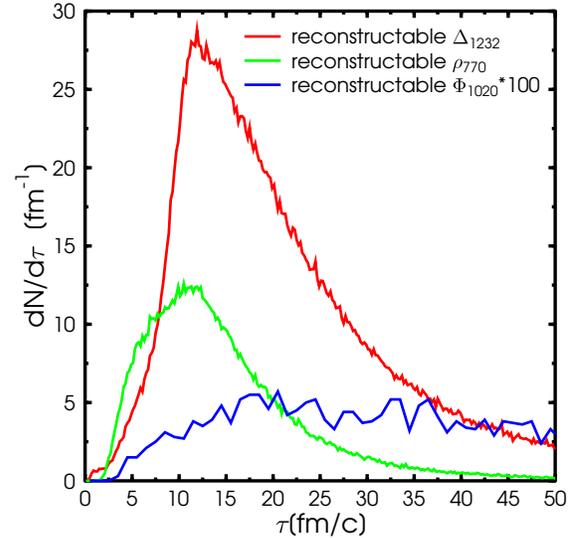,width=.5\textwidth,angle=0}} \caption{
Decay time distribution of reconstructable resonances for  Pb+Pb
collisions at 158~AGeV. Detectable $\Delta_{1232}$ and $\rho_{770}$
resonances stem from late stages of the collision. The $\Phi$ mesons
decays mostly outside of the collision zone (Note that the $\Phi$
distribution is multiplied by 100). } \label{f4}
\end{figure}

Fig. \ref{f4} shows the decay time distribution of reconstructable
resonances in Pb+Pb at 158 AGeV. Most reconstructable
$\Delta^0_{1232}$'s and $\rho^0_{770}$'s stem from the late stage of
the collision ($t_{\rm decay}\ge 10$~fm/c). The $\Phi_{1020}$ shows a
different behaviour, since it has a very small hadronic cross section
and a lifetime of 44 fm/c, which leads to a decay of a large part of
the $\Phi$ mesons outside the fireball.

\vspace{-25pt}

\begin{figure}[hbt]
\centerline{\epsfig {file=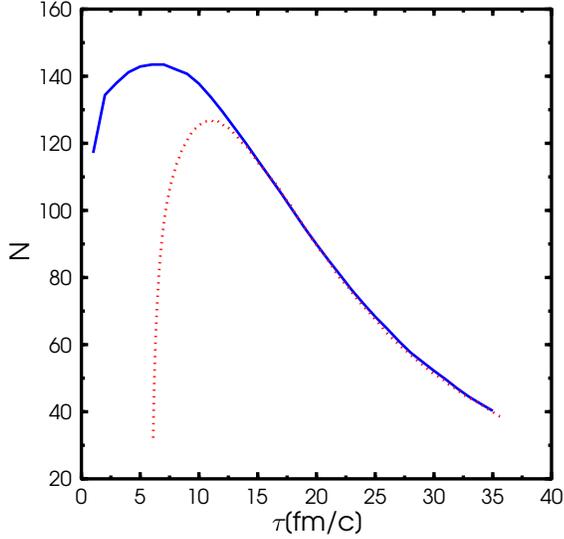,width=.5\textwidth,angle=0}}
\caption{ Time evolution of the $\Delta$ yield for Pb+Pb collision
at 158~AGeV. The full line is the evolution in the standard
calculation. The dashed line depicts the evolution of the
$\Delta_{1232}$ when all resonances were forced to decay at
6~fm/c.}\label{f5}
\end{figure}

In order to investigate the refeeding aspects two scenarios are
studied. The first scenario allows for a qualitative understanding
of the refeeding process: Here, the calculation is stopped roughly
at the chemical freeze-out at 6 fm/c \cite{Bleicher:2002dm}. At this
point all resonances are forced to decay. Then the calculation is
continued. The full line in Fig. \ref{f5} shows the time evolution
of all $\Delta_{1232}$ resonances since the beginning. The dotted
line shows the time evolution of the $\Delta_{1232}$ resonances
which regenerate after the complete decay at 6 fm/c. On clearly
observes that the regeneration of the $\Delta$ resonances is strong
and allows the delta yield to regain its full strength after
additional 6~fm/c propagation.

\begin{figure}[hbt]
\centerline{\epsfig {file=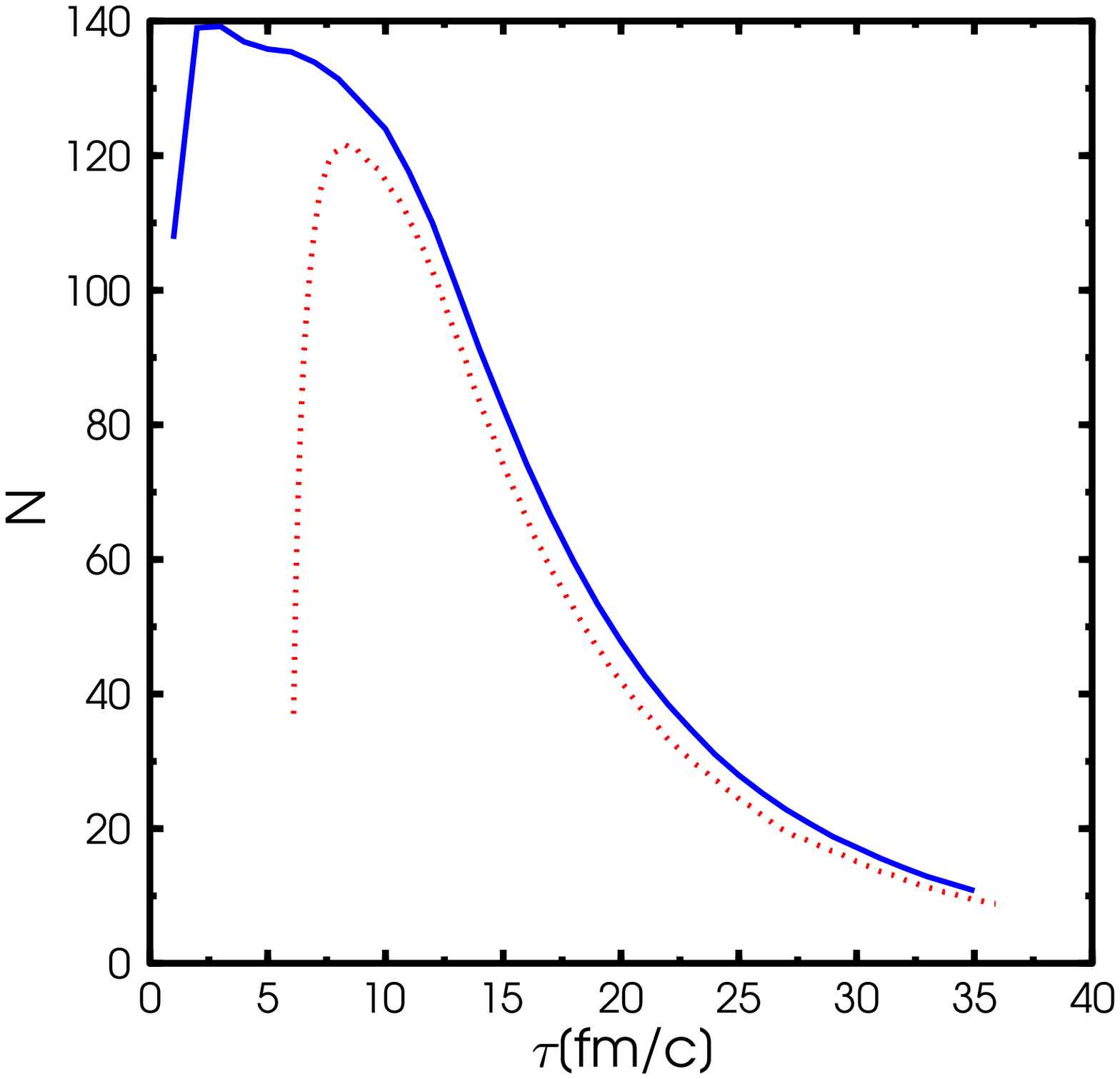,width=.5\textwidth,angle=0}}
\caption{Time evolution of the $\rho$ yield for Pb+Pb collision at
158~AGeV. The full line is the evolution in the standard
calculation. The dashed line depicts the evolution of the $\rho$
when all resonances were forced to decay at 6~fm/c. } \label{f6}
\end{figure}

Let us now turn to the $\rho$ resonance as shown in  Figure
\ref{f6}. The $\rho_{770}$ does not reach its initial value as the
$\Delta_{1232}$ does. This is because of the smaller and strongly
energy dependent cross sections for the reaction $\pi\pi \rightarrow
\rho_{770}$ compared to the reaction $p\pi \rightarrow
\Delta_{1232}$. While for a break-up time of 6~fm/c regeneration of
the rho is still possible, absorption effects take over  when the
system is more cooled down because most of the $\pi\pi$ interactions
will be at $\sqrt s_{\pi\pi}\ll m_\rho$.

\vspace{-25pt}

\vspace{-25pt}

\vspace{-25pt}

\begin{figure}[hbt]
\centerline{\epsfig {file=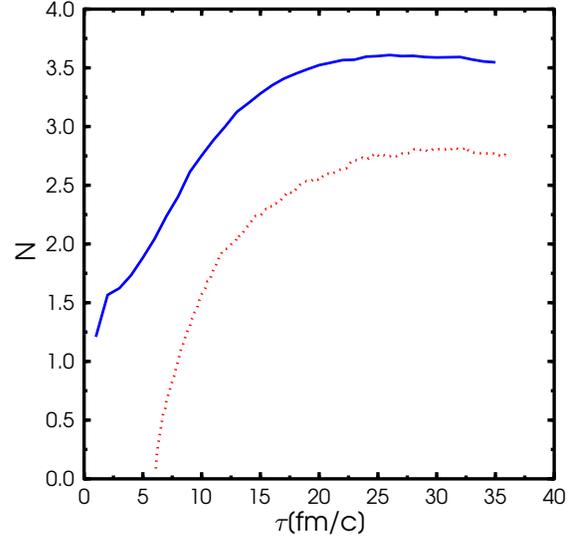,width=.5\textwidth,angle=0}}
\caption{Time evolution of the $\Phi$ yield for Pb+Pb collision at
158~AGeV. The full line is the evolution in the standard
calculation. The dashed line depicts the evolution of the $\Phi$
when all resonances were forced to decay at 6~fm/c. } \label{f7}
\end{figure}

Let us now probe this assumption with a particle with more exotic
decay branchings. Figure \ref{f7} depicts the number of $\Phi$
resonances as a function of time - again for the full calculation
(full line) and the modified calculation (dashed line). The
multiplicity of the $\Phi$ meson resonance does not reach the value
from the full calculation, but falls short by 40\% even after
additional 30~fm/c. In contrast to the $\rho$ and $\Delta$
regeneration is difficult in the $\Phi$ channel, because the
probability for $K\overline K\rightarrow\Phi$ is suppressed due to
the low Kaon densities and the small width of the $\Phi$.

\begin{figure}[hbt]
\centerline{\epsfig
{file=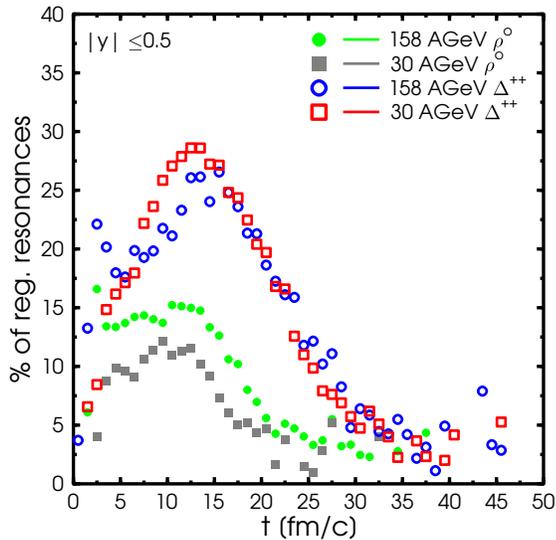,width=.5\textwidth,angle=0}} \caption{ Ratio
of regenerated/all decayed resonances as a function of time for
Pb+Pb collisions at 30~AGeV and 158~AGeV. Open symbols indicate the
$\Delta$ resonance, while full symbols show the $\rho$ resonance. }
\label{f8}
\end{figure}

Let us now turn to a quantitative analysis of the regeneration
processes as shown in Fig. \ref{f8}.  Here, we track all decay
products of a resonance. When a daughter particle from a resonance
decay directly reforms a resonance of the same type in the next
interaction it is marked as 'regenerated'. Note that only
$\rho^0_{770}$ into $\rho^0_{770}$ or $\Delta^{++}_{1232}$ into
$\Delta^{++}_{1232}$ are counted as directly 'regenerated'. This
explains the low percentage of regenerated resonances in the early
stage of the collision, where resonances are directly produced, e.g.
from $pp\rightarrow \Delta + X$.

The Delta (open symbols) regeneration peaks around 10-12~fm/c with a
direct regeneration probability between 25\% (158~AGeV) and 30\%
(30~AGeV) and stays strong throughout the final expansion stage of
the reaction. In contrast, the direct rho (full symbols)
regeneration is weak (only 10-15\% at both energies) and becomes
negligible after 12~fm/c.

To summarize, the importance of the regeneration of resonances in
heavy ion collisions was studied. It was shown that a full
regeneration of the $\Delta_{1232}$ abundancies on time scales of
6~fm/c is possible within the present model.  In contrast to the
rho, a large regeneration probability of Deltas was found  which
might explain the slight enhancement of the $\Delta^{++}/p$ ratio
with centrality. Further investigation of the regeneration and
absorption dynamics are currently under way.

\section*{Acknowledgements}

This work was supported by  GSI, DFG and BMBF. The computational
resources  have been provided by the Center for Scientific Computing
at Frankfurt. S.V. thanks the organizers for the great time at
Winter Meeting in Bormio.


\end{document}